\documentclass[letterpaper]{article}

\usepackage{isea}
\usepackage[numbers]{natbib}
\usepackage[pdftex]{graphicx}           
\usepackage{times}
\usepackage{helvet}
\usepackage{courier}

\usepackage[caption=false]{subfig}
\usepackage{url}                  
\usepackage[T1]{fontenc}
\usepackage{nicefrac}             
\usepackage{indentfirst}          
\usepackage[super, negative]{nth}
\usepackage{amsfonts}
\usepackage{amssymb}
\usepackage{amsmath}
\usepackage{multirow}
\usepackage{balance}
\usepackage{subfloat}
\usepackage{bytefield}
\usepackage{mathdots}
\usepackage{tikz}
\usetikzlibrary{positioning,decorations.pathreplacing,calc,fit}
\newcommand\tikzmark[1]{%
  \tikz[remember picture] \node (#1) {};
}

\usepackage{xspace}
\usepackage{array,booktabs}
\usepackage{latexsym}
\usepackage{pifont}
\usepackage{colortbl}

\usepackage{algorithm}
\usepackage[noend]{algpseudocode}
\usepackage{listings}
\usepackage{boldline}

\usepackage{xcolor}

\definecolor{MACRouge}{RGB}{255,38,0}
\definecolor{MACBleu}{RGB}{4,51,255}
\definecolor{MACVert}{RGB}{0,143,0}
\definecolor{MACOrange}{RGB}{255,147,0}
\definecolor{MACMagenta}{RGB}{255,64,255}
\definecolor{MACAsperge}{RGB}{146,144,0}
\definecolor{MACGris}{RGB}{169,169,169}
\definecolor{MACBordeau}{RGB}{148,23,81}
\definecolor{light-gray}{gray}{0.95}

\newcommand{\dfn}[1]{\textit{#1}}            

\newcommand{\ipsix}{\textsc{Ipv6}\xspace}

\newcommand{\ingress}{\textsc{Ingress}\xspace}
\newcommand{\egress}{\textsc{Egress}\xspace}

\newcommand{\ioam}{\textsc{Ioam}\xspace}
\newcommand{\ioamLong}{\textbf{I}n Situ \textbf{O}perations, \textbf{A}dministration, and \textbf{M}aintenance (\ioam)\xspace}

\newcommand{\lwt}{\textsc{Lwt}\xspace}
\newcommand{\lwts}{\textsc{Lwt}s\xspace}

\newcommand{\dut}{\textsc{DUT}\xspace}
\newcommand{\sr}{\textsc{Sr}v6\xspace}
\newcommand{\srmpls}{\textsc{Sr-Mpls}\xspace}
\newcommand{\srLong}{\textbf{S}egment \textbf{R}outing over \ipsix (\sr)\xspace}
\newcommand{\rpl}{\textsc{Rpl}\xspace}
\newcommand{\rplLong}{\textbf{R}outing \textbf{P}rotocol for \textbf{L}ow-Power and Lossy Networks (\rpl)\xspace}
\newcommand{\mplsLong}{\textbf{M}ulti\textbf{P}rotocol \textbf{L}abel \textbf{S}witching (\textsc{Mpls})\xspace}
\newcommand{\mpls}{\textsc{Mpls}\xspace}
\newcommand{\isp}{ISP\xspace}
\newcommand{\nsh}{\textsc{Nsh}\xspace}
\newcommand{\nshLong}{\textbf{N}etwork \textbf{S}ervice \textbf{H}eader (\nsh)\xspace}
\newcommand{\ipsixeh}{Extension Header\xspace}
\newcommand{\ipsixhbh}{Hop-by-Hop\xspace}
\newcommand{\pto}{\textsc{Pto}\xspace}
\newcommand{\ptos}{\textsc{Pto}s\xspace}
\newcommand{\ptoLong}{\textbf{P}re-allocated \textbf{T}race \textbf{O}ption (\pto)\xspace}
\newcommand{\sid}{\textsc{Sid}\xspace}
\newcommand{\sids}{\textsc{Sid}s\xspace}
\newcommand{\cpu}{\textsc{CPU}\xspace}
\newcommand{\cpus}{\textsc{CPUs}\xspace}

\makeatletter
\newcommand{\DeclareLatinAbbrev}[2]{%
  \DeclareRobustCommand{#1}{%
    \@ifnextchar{.}{\emph{#2}}{%
      \@ifnextchar{,}{\emph{#2.}}{%
        \@ifnextchar{!}{\emph{#2.}}{%
          \@ifnextchar{?}{\emph{#2.}}{%
            \@ifnextchar{)}{\emph{#2.}}{%
              {\emph{#2.\ }}}}}}}}%
}
\makeatother
\DeclareLatinAbbrev{\eg}{e.g}
\DeclareLatinAbbrev{\Eg}{E.g}
\DeclareLatinAbbrev{\ie}{i.e}
\DeclareLatinAbbrev{\Ie}{I.e}
\DeclareLatinAbbrev{\etc}{etc}

\lstdefinestyle{block}{
  language=C,
  tabsize=4,
  frame=single,
  basicstyle=\ttfamily\scriptsize,
  keywordstyle=\color{blue},
  commentstyle=\color{gray}\itshape,
  stringstyle=\color{teal},
  otherkeywords={__u32,u16,u8,...},
  breaklines=true,
  float,
  columns=fullflexible,
  escapeinside={(*}{*)},
  captionpos=b,
}

\graphicspath{{.}{./Pictures/}}

\title{Mitigating the Double-Reallocation Issue for IPv6 \\ Lightweight Tunnel Encapsulations in the Linux Kernel}
\author{Justin Iurman, Emilien Wansart, Maxime Goffart, Benoit Donnet\\
Universit\'e de Li\`ege -- Montefiore Institute\\
Li\`ege, Belgium\\
\url{{firstname.lastname}@uliege.be}\\
\newline
\newline
}
\begin{document}
\maketitle

\begin{abstract}
  Lightweight Tunnels (\lwts) in the Linux kernel enable efficient per-route tunneling and are widely used by protocols such as \ioamLong, \srLong, and \rplLong. However, a performance issue was detected in their implementations, where a double-reallocation of socket buffers occurs under specific conditions, leading to significant throughput degradation. This paper investigates the root cause of the issue, which depends on the architecture of the Central Processing Unit (\cpu) and the Network Interface Card (NIC). We propose a patch for the Linux kernel to fix this problem, replacing the double-reallocation with a single, efficient one. Performance evaluation demonstrates that the patch eliminates the inefficiency, improving forwarding rates by up to 28.8\% for affected protocols.
\end{abstract}

\section{Keywords}
Linux, Networking, Performance, \ioam, \sr, \rpl

\section{Introduction}\label{intro}
In the Linux kernel, \textsc{Ip}\xspace tunnels are commonly implemented using Lightweight Tunnels (\lwts)~\cite{lwts}. \lwts enable the creation of tunnels on a per-route basis, allowing custom encapsulation mechanisms to be defined through user-specified functions based on the packet destination. Various kernel networking units leverage \lwts, including:
\begin{itemize}
  \item \ioamLong~\cite{rfc9197}, a network telemetry protocol carrying operational data from devices (\eg, routers or switches), such as latency and buffer size, directly into packet headers as they traverse the network.
  \item \srLong~\cite{rfc8402}, a source routing protocol encoding the routing path into \ipsix headers, enabling simplified traffic engineering and flexible network programmability.
  \item \rplLong~\cite{rfc6550}, a routing protocol designed for resource-constrained networks, particularly in Internet-of-Things (IoT) building optimized multi-hop tree-like structures for reliable communication in dynamic and lossy environments.
  \item \mplsLong~\cite{rfc3031}, a technology designed to speedup forwarding decisions (through exact label matching instead of longest prefix matching on \textsc{Ip}\xspace addresses) but is nowadays mainly deployed for providing IGP/BGP scalability and virtual private network (VPN) services~\cite{rfc2917}.
  \item \textbf{I}dentifier \textbf{L}ocator \textbf{A}ddressing for \ipsix (\textsc{Ila})~\cite{herbert-intarea-ila-01}, a way to differentiate between location and identity of a network node.
  \item Virtual \texttt{xfrm} interfaces~\cite{xfrm} used for route-based VPN tunnels.
\end{itemize}

When running some \ioam performance evaluations, we detected an issue in its implementation. The problem lied in the \lwt encapsulation of \ioam and occurs exclusively in this mode of operation. Such a behavior was particularly hard to understand since it occurred only under some conditions, that depend on the architecture of the Central Processing Unit (\cpu) and the Network Interface Card (NIC). For those specific cases, and during the \lwt encapsulation in packets, the kernel performs a double-reallocation of socket buffers instead of a single one, leading to obvious performance degradation. After some investigations, we found similar code patterns causing the exact same issue in the \sr and \rpl implementations. To quantify the impact on performance, we measured the packet forwarding rate on a recent kernel version and observed up to 28.8\% degradation.

This paper aims at addressing the aforementioned issue by proposing a Linux kernel patch that eliminates this inefficiency.  Our patch replaces the double-reallocation with a single, optimized one.  In addition, to validate its effectiveness, we conduct a performance evaluation of the three affected protocols, comparing the original implementation with our patched version, demonstrating its ability to improve forwarding efficiency.

\section{Background}\label{background}
This paper focuses on a bug in the Lightweight Tunnel (\lwt) implementation of \ioam, \sr, and \rpl protocols.  In this section, we provide some background to readers by describing the aforementioned protocols, the structure of a socket buffer, and the role of the \texttt{skb\_cow\_head()} function in the Linux kernel.

\subsection{\ioamLong}

\begin{figure}[!t]
  \begin{center}
    \includegraphics[width=\linewidth]{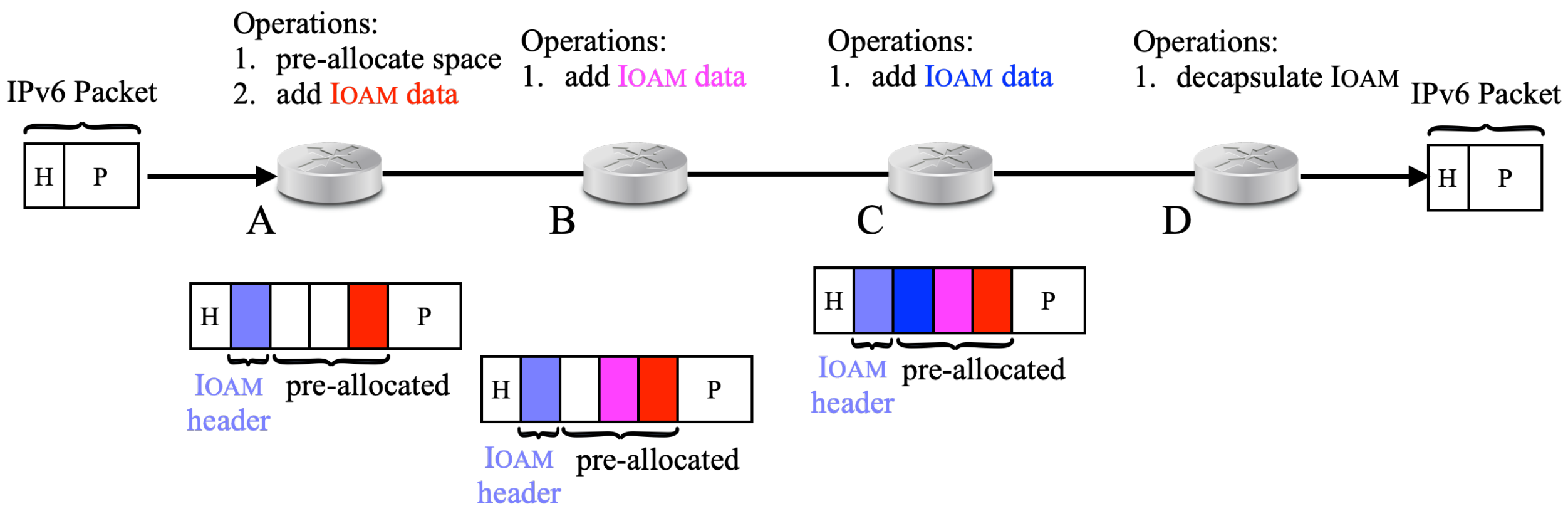}
  \end{center}
  \caption{Generic example of \ioam data insertion. ``H'' corresponds to the \ipsix header, while ''P`` is the \ipsix payload.}
        \label{back:ioam.fig}
\end{figure}

\begin{figure}[!t]
	\begin{center}
		\begin{bytefield}{32}
			\bitbox{10}{\ipsix Header} & \bitbox{12}{\ioam Option}&\bitbox{10}{\ipsix Payload}\\
			\bitbox[]{8}{}&\bitbox[]{1}{\tikzmark{a}}&\bitbox[]{13}{}&\bitbox[]{2}{\tikzmark{b}}\\[1ex]
			\bitbox[]{1}{\tikzmark{c}}&\bitbox[]{30}{}&\bitbox[]{1}{\tikzmark{d}}\\
			\bitheader{0,7,15,23,31}\\
			\bitbox{8}{Next Header}\bitbox{8}{Hdr Ext Len}\bitbox{16}{Padding}\\
			\bitbox{8}{Option-Type}\bitbox{8}{Opt Data Len}\bitbox{8}{Reserved}\bitbox{8}{\small IOAM Opt-Type}\\
			\bitbox{16}{Namespace-ID}\bitbox{5}{NodeLen}\bitbox{4}{Flags}\bitbox{7}{\small RemainingLen}\\
			\bitbox{24}{IOAM Trace-Type}\bitbox{8}{Reserved}\\
			\bitbox{32}{node data list [0..n]}
		\end{bytefield}
	\end{center}
	\vspace{-0.5cm}
	\caption{\ipsixhbh with an \ioam \pto Option~\cite{rfc9197}.}
	\label{back:ioamh.fig}
\end{figure}
\tikz[remember picture,overlay] \draw[-, line width=0.35mm] ([xshift=0.4cm]a.north) -- node[above]{} (c.south);
\tikz[remember picture,overlay] \draw[-, line width=0.35mm] ([xshift=-0.25cm]b.north) -- node[above]{} ([xshift=0.14cm]d.south);

\ioamLong~\cite{rfc9197} provides a way to collect telemetry data from network devices (\eg, routers or switches) along a path inside a given \ioam domain (e.g., within an \isp or Data Center Networks). ``In Situ'' refers to the fact that \ioam does not use dedicated packets for carrying the telemetry data. Rather, it relies on existing user traffic. \ioam telemetry can be encapsulated in a variety of protocols, including \ipsix~\cite{rfc9486} or \nshLong~\cite{rfc9452}. In this paper, we only consider the \ipsix encapsulation.

An \ioam domain contains three types of nodes. First, the \dfn{encapsulating} nodes, located at the entry points (or \ingress) of the domain, are responsible for appending the \ipsix \ipsixeh required for storing \ioam telemetry (node $A$ in Fig.~\ref{back:ioam.fig}). Second, the \dfn{transit} nodes add telemetry data inside the existing \ipsix \ipsixeh (nodes $B$ and $C$ in Fig.~\ref{back:ioam.fig}). Finally, the \dfn{decapsulating} nodes, positioned at the exit points (or \egress) of the domain, are responsible for removing the \ipsix \ipsixeh (node $D$ in Fig.~\ref{back:ioam.fig}). The format of the \ipsix \ipsixeh to carry \ioam telemetry is represented in Fig.~\ref{back:ioamh.fig}.

There are two possibilities for encapsulating the \ioam data in \ipsix.  If the packet source is the encapsulating node, the \ioam data fields can be embedded in the existing \ipsix header. Otherwise, the \ioam data must be added inside an additional \ipsix header, since in-transit modification of existing headers is forbidden~\cite{rfc8200}. This approach leads to the creation of an \ipsix-in-\ipsix tunnel across the \ioam domain. \ioam telemetry inside \ipsix can be encapsulated in either a \ipsixhbh \ipsixeh, which is processed by every node, or a Destination \ipsixeh, which is only processed by the destination~\cite{rfc9486}.

While \ioam defines several possibilities for adding telemetry data to packets~\cite{rfc9197,rfc9326}, this paper only considers the \ptoLong in which encapsulating nodes allocate the required space inside the \ipsix \ipsixeh in advance. \pto is illustrated in Fig.~\ref{back:ioam.fig}, where node $A$ pre-allocates the required space for adding \ioam data.

In the kernel, the following encapsulation modes are supported for \ipsix \ioam:
\begin{itemize}
  \item \dfn{Inline} mode: Directly added into the original packet header, which allows telemetry data to be included without creating an additional encapsulating header.
  \item \dfn{Encapsulation} (or \dfn{Tunnel}) mode: The original packet is encapsulated within a new \ipsix outer header, which contains telemetry data. This creates an \ipsix-in-\ipsix tunnel.
  \item \dfn{Automatic} mode: The kernel either applies the \dfn{Inline} or \dfn{Tunnel} mode depending on the source of packets (\ie, \dfn{Inline} mode if it is the source, \dfn{Tunnel} mode otherwise).
\end{itemize}

\subsection{\srLong}

\begin{figure}[!t]
  \begin{center}
    \begin{bytefield}{32}
        \bitbox{10}{\ipsix Header} & \bitbox{12}{\sr Header}&\bitbox{10}{\ipsix Payload}\\
        \bitbox[]{8}{}&\bitbox[]{1}{\tikzmark{a}}&\bitbox[]{13}{}&\bitbox[]{2}{\tikzmark{b}}\\[1ex]
        \bitbox[]{1}{\tikzmark{c}}&\bitbox[]{30}{}&\bitbox[]{1}{\tikzmark{d}}\\
        \bitheader{0,7,15,23,31}\\
        \bitbox{8}{Next Header}\bitbox{8}{Hdr Ext Len}\bitbox{8}{Routing Type}\bitbox{8}{Segments Left}\\
        \bitbox{8}{Last Entry}\bitbox{8}{Flags}\bitbox{16}{Tag}\\
        \bitbox{32}{Segment List[0] (128-bit \ipsix address)}\\
        \bitbox{32}{Segment List[1] (128-bit \ipsix address)}\\
        \bitbox{32}{\ldots}\\
        \bitbox{32}{Segment List[n] (128-bit \ipsix address)}\\
        \bitbox{32}{Optional TLVs (variable size)}\\
      \end{bytefield}
  \end{center}
\vspace{-0.5cm}
  \caption{\sr Header~\cite{rfc8754}.}
  \label{back:srv6.fig}
\end{figure}
\tikz[remember picture,overlay] \draw[-, line width=0.35mm] ([xshift=0.4cm]a.north) -- node[above]{} (c.south);
\tikz[remember picture,overlay] \draw[-, line width=0.35mm] ([xshift=-0.25cm]b.north) -- node[above]{} ([xshift=0.14cm]d.south);

In a nutshell, Segment Routing~\cite{rfc8402} is a loose source routing paradigm based on an ordered list of \dfn{segments} (\ie, one or more instructions). Each segment can enforce a topological requirement (\eg, pass through a node or an interface) or a service requirement (\eg, execute an operation on the packet). Over the years, Segment Routing has found a suitable usage for many use cases such as network monitoring, traffic engineering, or failure recovery~\cite{sr-survey}, among others. Two forwarding plane implementations are proposed for Segment Routing: Segment Routing over \mpls -- \srmpls~\cite{sr_mpls} and Segment Routing over \ipsix -- \sr~\cite{rfc8986}.  \srmpls requires no change to the \mpls forwarding plane, while \sr is based on a Routing Header called \sr Header. In this paper, we only consider \sr.

Segment Routing defines multiple types of segments, but the two most common are \dfn{node segments} and \dfn{adjacency segments}. A \dfn{node segment} represents the IGP least cost path between any router and a specified prefix. These segments can contain one or multiple IGP hops and have domain-wide significance. In normal Segment Routing operations, every Segment Routing router will announce a \dfn{node segment} for itself, allowing any router in the domain to know how to reach it. An \dfn{adjacency segment} represents an IGP adjacency between two routers and will cause a packet to traverse that specified link. These segments only have local significance. In normal Segment Routing operations, every Segment Routing router advertises an \dfn{adjacency segment} for each of its links.

Each segment is identified by a unique number, a \dfn{Segment IDentifier} (\sid) implemented as a 128-bit \ipsix address in \sr, enabling deployments over non-\mpls networks or areas without \mpls, such as data centers. This implementation simplifies deployments as it only requires advertising \ipsix prefixes. \sids are encoded within the Routing Extension Header known as \sr Header~\cite{rfc8754} (see Fig.~\ref{back:srv6.fig}).

In the fashion of \ioam, the Linux kernel supports both the \dfn{Inline} and \dfn{Tunnel} encapsulation modes for \sr. Additionally, \sr features a reduced encapsulation variant (referred to as ``Red'') to optimize header size.  It also offers a Layer-2 (``L2'') encapsulation variant, where the received frame is encapsulated within the \ipsix packet.

\subsection{\rplLong}

\begin{figure}[!t]
	\begin{center}
		\begin{bytefield}{32}
			\bitbox{10}{\ipsix Header} & \bitbox{12}{\rpl Header}&\bitbox{10}{\ipsix Payload}\\
			\bitbox[]{8}{}&\bitbox[]{1}{\tikzmark{a}}&\bitbox[]{13}{}&\bitbox[]{2}{\tikzmark{b}}\\[1ex]
			\bitbox[]{1}{\tikzmark{c}}&\bitbox[]{30}{}&\bitbox[]{1}{\tikzmark{d}}\\
			\bitheader{0,7,15,23,31}\\
			\bitbox{8}{Next Header}\bitbox{8}{Hdr Ext Len}\bitbox{8}{Routing Type}\bitbox{8}{Segments Left}\\
			\bitbox{4}{CmprI}\bitbox{4}{CmprE}\bitbox{4}{Pad}\bitbox{20}{Reserved}\\
			\bitbox{32}{Addresses[1...n]}\\
		\end{bytefield}
	\end{center}
	\vspace{-0.5cm}
	\caption{\rpl Header~\cite{rfc6554}.}
	\label{back:rpl.fig}
\end{figure}
\tikz[remember picture,overlay] \draw[-, line width=0.35mm] ([xshift=0.4cm]a.north) -- node[above]{} (c.south);
\tikz[remember picture,overlay] \draw[-, line width=0.35mm] ([xshift=-0.25cm]b.north) -- node[above]{} ([xshift=0.14cm]d.south);

\rpl~\cite{rfc6550} is a routing protocol for wireless networks, especially suitable for resources-constrained networks such as Internet-of-Things (IoT). It is a Distance Vector Routing Protocol that creates a tree-like routing topology called the Destination Oriented Directed Acyclic Graph (DODAG), rooted towards one or more nodes called the root node or sink node.

For downward routing in non-storing mode (one of \rpl's modes of operation), \rpl uses a Source Routing header to deliver datagrams, as shown in Fig.~\ref{back:rpl.fig}. Just like \sr, this header contains a list of addresses that the packet must traverse to reach its destination. It comes with a compression mechanism, called Address Abbreviation, designed to reduce the size of the Routing header. If all nodes in the path share a common prefix, only the unique interface identifiers of each node are included. Therefore, the \rpl header introduces the \texttt{CmprI}, \texttt{CmprE}, and \texttt{Pad} fields to allow compaction of the \texttt{Addresses[1...n]} vector when all entries share the same prefix as the \ipsix destination address of the packet. The \texttt{CmprI} and \texttt{CmprE} fields indicate the number of prefix octets that are shared with the \ipsix destination address of the packet. The shared prefix octets are not carried within the \rpl header. The \texttt{Pad} field indicates the number of unused octets that are used for padding.

On the contrary to \ioam and \sr, the Linux kernel only supports the \dfn{Inline} encapsulation mode for \rpl.

\subsection{Linux Kernel Socket Buffer}
A socket buffer (\texttt{sk\_buff}) in the Linux kernel is a core networking structure used to represent network packets. It acts as a container for the packet's metadata, enabling efficient processing, routing, and management of network traffic.

\begin{figure}[!t]
  \begin{center}
    \includegraphics[width=\linewidth]{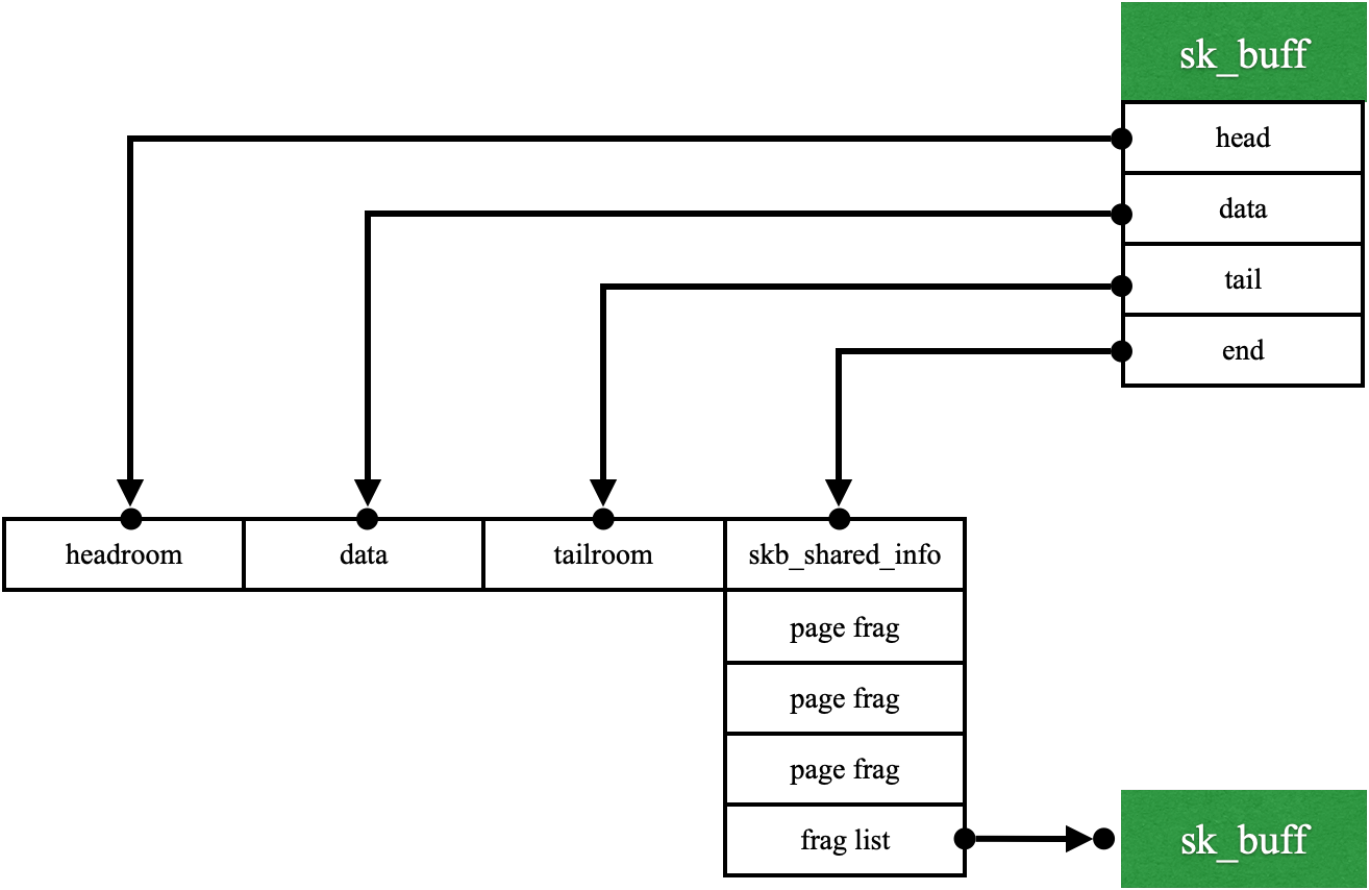}
  \end{center}
  \caption{Basic \texttt{sk\_buff} diagram~\cite{skbuff}.}
  \label{lst:skbuff}
\end{figure}

Fig.~\ref{lst:skbuff} illustrates the layout of the \texttt{sk\_buff} structure.  It consists of several key components:
\begin{itemize}
  \item \texttt{headroom}: Free space available for prepending headers.
  \item \texttt{data}: This area holds the headers and payload.
  \item \texttt{tailroom}: Free space available for adding trailing data if necessary.
  \item \texttt{skb\_shared\_info}: This structure holds an array of pointers to read-only data.
\end{itemize}

This design allows the kernel to modify headers or add encapsulations while minimizing memory reallocations.

\subsection{The \texttt{skb\_cow\_head()} function}
Listing~\ref{lst:sch} shows the definition of the \texttt{skb\_cow\_head()} function. Its purpose is to make sure that a socket buffer (\ie, \texttt{skb} argument) has at least the desired headroom size (\ie, \texttt{headroom} argument). It is usually used when one only needs to add headers but does not need to modify the data. This function does nothing when there is enough headroom. Otherwise, it reallocates more space to accommodate for the required headroom. In that case, the desired headroom size (in bytes) is aligned to the next multiple of a specific value which is totally dependent on the architecture of the \cpu and represents its cache line size. In the Linux kernel, the minimum allowed size for a cache line is $32$ bytes and can typically be either $64$, $128$ or even $256$ bytes depending on the \cpu.

As an example and in order to illustrate, let us assume we want to add $112$ bytes to the header of a socket buffer (\texttt{skb}), whose current headroom size is $64$. Let us also assume a \cpu with a $32$-byte cache line size. After a call to \dfn{skb\_cow\_head(skb, 112)}, the new headroom size would be $128$. The reasoning is the following: ($i$) the \cpu cache line size is added to the current headroom size ($64 + 32 = 96$), which is smaller than $112$ bytes; ($ii$) the \cpu cache line size is re-added to that value ($96 + 32 = 128$), which is now bigger or equal to $112$ bytes. Note that after the insertion of the $112$ bytes in the header, the new headroom size would be $16$ bytes ($128 - 112$).

\begin{lstlisting}[caption=Definition of the \texttt{skb\_cow\_head()} function.,label={lst:sch},style=block,language=C]
static inline int skb_cow_head(
	struct sk_buff *skb,
	unsigned int headroom
);
\end{lstlisting}

\section{Double-Reallocation Issue}\label{issue}
This paper investigates an issue in the \lwt implementation of \ioam, \sr, and \rpl protocols. In normal cases, when an extra header is added to an existing packet, the kernel triggers a reallocation of the socket buffer if its headroom cannot accommodate it. However, in some cases and under specific conditions (see below), the kernel triggers two consecutive reallocations, resulting in performance degradation.

\begin{lstlisting}[caption=Simplified code pattern that triggers a double-reallocation of socket buffers in some cases., label={lst:pattern},style=block,language=C,numbers=left,escapechar=|]
	err = skb_cow_head(skb, hdrlen + skb->mac_len);|\label{lst:pattern:1}|
	// add the new header ("hdrlen" bytes)|\label{lst:pattern:2}|
	// get "dst" in cache or resolve the new "dst" |\label{lst:pattern:3}|
	err = skb_cow_head(skb, LL_RESERVED_SPACE(dst->dev));|\label{lst:pattern:4}|
\end{lstlisting}

Listing~\ref{lst:pattern} shows the problematic code pattern that is common to the \lwt implementation of \ioam, \sr, and \rpl protocols. First, line~\ref{lst:pattern:1} ensures sufficient headroom in the socket buffer (\dfn{skb}) before adding a new header (\dfn{hdrlen} bytes), and anticipates the Layer-2 header reconstruction (\dfn{mac\_len} bytes, \eg, $14$ for Ethernet). Then, line~\ref{lst:pattern:2} inserts the new header in the socket buffer, while line~\ref{lst:pattern:3} gets the destination entry (\dfn{dst}) stored in the cache and, if empty, resolves the new destination entry based on the new packet header. This is because the original output device may no longer be the correct one (\eg, an \ipsix-\ipsix tunnel encapsulation with different source or destination addresses than the original packet, leading to a different egress interface). Finally, line~\ref{lst:pattern:4} ensures sufficient headroom in the socket buffer (\dfn{skb}) to accommodate the maximum hardware header length of the output device (\dfn{dst->dev}), including any extra headroom the NIC may need. The total length returned by the macro \dfn{LL\_RESERVED\_SPACE()} is aligned to 16-byte multiples for machine alignment needs. Indeed, \cpus often take a performance hit when accessing unaligned memory locations. For instance, since an Ethernet header is $14$ bytes long, network drivers often end up with the \ipsix header at an unaligned offset. The \ipsix header can be aligned by shifting the start of the packet by $2$ bytes.

To illustrate a case that would trigger a double-reallocation based on Listing~\ref{lst:pattern}, let us assume we want to add an extra header ($40$ bytes) in the socket buffer, whose current headroom size is $22$. Let us also assume a \cpu with a $32$-byte cache line size, and Ethernet as Layer-2. First, in line~\ref{lst:pattern:1}, \dfn{hdrlen} is $40$ (the extra header size) and \dfn{mac\_len} is $14$ (Ethernet header size). After that, the headroom size becomes $54$ (\ie, $22 + 32$, the old headroom size plus the cache line size). Then, in line~\ref{lst:pattern:2}, we insert the extra $40$-byte header. Therefore, the headroom size is now $14$ (\ie, $54 - 40$, the old headroom size minus the extra header size). Finally, in line~\ref{lst:pattern:4}, we ensure that the headroom can accommodate the hardware header length (\ie, $16$ bytes in this case). Since the current headroom size is too small ($14 < 16$), a second reallocation is performed. As a result, the headroom size becomes $46$ (i.e., $14 + 32$, the old headroom size plus the cache line size). The root cause of the double-reallocation comes from the difference between line~\ref{lst:pattern:1} and line~\ref{lst:pattern:4} in listing~\ref{lst:pattern}, where \dfn{mac\_len} is too generic and often smaller than \dfn{LL\_RESERVED\_SPACE()} due to alignment requirements. A solution to avoid this problem would be to use the latter in both lines, instead of \dfn{mac\_len}. However, as already mentioned, the output device in line~\ref{lst:pattern:1} may not be the same after line~\ref{lst:pattern:3}. Therefore, this solution cannot work as is.

{\setlength{\doublerulesep}{0pt}
	\setlength{\tabcolsep}{5pt}
	\begin{table}[!t]
		\small
		\begin{center}
			\resizebox{\columnwidth}{!}{%
				\begin{tabular}{m{.35 \columnwidth}m{.55 \columnwidth}}
					\hline\hline\hline\hline
					\multicolumn{2}{c}{\textbf{\ioam}}\\
					\hline
					Inline mode & \pto of 236 or 240 bytes\\
					Encap. mode & \pto of 196 or 200 bytes\\
					\hline
					\multicolumn{2}{c}{\textbf{\sr}}\\
					\hline
					Inline mode & None\\
					Encap. mode & For 13, 17, 21, 25, 29, 33, .. segments\\
					Encap. L2 mode & For 13, 17, 21, 25, 29, 33, .. segments\\
					Encap. Red mode & For 14, 18, 22, 26, 30, 34, .. segments\\
					Encap. L2 Red mode & For 14, 18, 22, 26, 30, 34, .. segments\\
					\hline
					\multicolumn{2}{c}{\textbf{\rpl}}\\
					\hline
					Inline mode & None\\
					\hline\hline\hline\hline
				\end{tabular}
			}
		\end{center}
		\caption{Cases triggering the double-reallocation for an x86 architecture and an Intel XL710 NIC.}
		\label{tbl:occur}
	\end{table}
}

As discussed, the double-reallocation mainly depends on the \cpu architecture (\ie, L1 cache line size) and, to a lesser extent, on the NIC (\ie, for alignment requirements and headroom allocation). Headroom allocation depends on the NIC. In our case, the network driver (\dfn{i40e}) allocates $192$ bytes for the headroom\footnote{With \dfn{legacy-rx} disabled by default. Otherwise, a cache line sized (\ie, $64$-byte for \dfn{x86}) headroom is allocated. In both cases, the double-reallocation happens because $192$ is a multiple of $64$.}. Therefore, before calling \dfn{skb\_cow\_head()} for the \lwt encapsulation, the headroom size of all socket buffers is $206$ bytes (\ie, $192 + 14$, because the data pointer was moved beyond the $14$-byte Ethernet header). Table~\ref{tbl:occur} summarizes the cases where a double-reallocation happens for the \lwt encapsulation of \ioam, \sr, and \rpl protocols. It is based on an \dfn{x86} architecture (\ie, $64$-byte cache line size), and an \texttt{Intel XL710} NIC (\dfn{i40e} driver).

Let us take \ioam in Table~\ref{tbl:occur} to illustrate the problem. All four cases produce the same overhead (\ie, \dfn{hdrlen} provided to \dfn{skb\_cow\_head()}), that is $256$ bytes for the entire \ipsixeh. Indeed, the Encap mode must include the extra \ipsix header (+$40$ bytes), which is identical to the Inline mode in terms of overhead. Moreover, a $236$-byte and a $240$-byte \pto both produce the same overhead, as padding is added to the \ipsixeh. If we apply the code in Listing~\ref{lst:pattern} to this case, we have: ($i$) an extra $256$-byte header to add and the $14$-byte Ethernet header to rebuild later, ($ii$) a $206$-byte headroom, and ($iii$) a $64$-byte cache line size. After line~\ref{lst:pattern:1}, the headroom becomes $270$ (\ie, $206 + 64$, the old headroom size plus the cache line size). Then, in line~\ref{lst:pattern:2}, we insert the extra $256$-byte header. Therefore, the headroom size is now $14$ (\ie, $270 - 256$, the old headroom size minus the extra header size). Finally, in line~\ref{lst:pattern:4}, we ensure that the headroom can accommodate the hardware header length (\ie, $16$ bytes in this case). Since the current headroom size is too small ($14 < 16$), a second reallocation is performed. As a result, the headroom size becomes $78$ (i.e., $14 + 64$, the old headroom size plus the cache line size).

Because the \ioam \pto is limited to a maximum of $244$ bytes, there are no repeated values\footnote{There would be more with \dfn{legacy-rx} enabled, depending on the cache line size.} with a double-reallocation like for \sr and \rpl (\ie maximum $127$ segments, meaning $2,040$ bytes). Nevertheless, the exact same logic applies based on the total overhead for each case, \eg, the total overhead of $21$ \sr segments (Encap mode) is $384$ bytes (\ie, extra \ipsix header plus \sr Routing header). Note that the double-reallocation does not happen for \sr with the Inline mode because it does not fall on the same boundaries anymore (\ie, -$40$ bytes compared to the Encap mode with the extra \ipsix header). On the other hand, \rpl (only the Inline mode is implemented in the kernel) comes with a compression mechanism, which makes it theoretically possible to trigger the double-reallocation issue. However, because \texttt{iproute2}~\cite{iproute2} has an input buffer limited to a maximum of $1,024$ characters, we could not add enough segments to make the total overhead large enough to trigger the issue.

Overall, it is possible to generalize the problem to other \cpu architectures: the smaller the cache line size, the more often the double-reallocation of socket buffers will occur. For instance, having a $32$-byte cache line would trigger the issue twice as frequently compared to a $64$-byte cache line. E.g., with a $32$-byte cache line, the double-reallocation would happen with \sr (Encap mode) for the following number of segments: $11$, $13$, $15$, $17$, $19$, $21$, $23$, $25$, etc (instead of $13$, $17$, $21$, $25$, etc with a $64$-byte cache line as in Table~\ref{tbl:occur}).

\section{Mitigation Solution}\label{mitigation}
\begin{lstlisting}[caption=Simplified code pattern to mitigate the double-reallocation of socket buffers.,label={lst:pattern2},style=block,language=C,numbers=left,escapechar=|]
	err = skb_cow_head(skb, hdrlen + dst_dev_overhead(cache_dst, skb));|\label{lst:pattern2:1}|
	// add the new header ("hdrlen" bytes)|\label{lst:pattern2:2}|
	// resolve the new "dst" if cache is empty|\label{lst:pattern2:3}|
	err = skb_cow_head(skb, LL_RESERVED_SPACE(dst->dev));|\label{lst:pattern2:4}|
\end{lstlisting}

\begin{lstlisting}[caption=Code of the new dst\_dev\_overhead() function.,label={lst:ddo},style=block,language=C,escapechar=|]
static inline unsigned int dst_dev_overhead(
		struct dst_entry *dst, struct sk_buff *skb)
{
	if (likely(dst))
		return LL_RESERVED_SPACE(dst->dev);

	return skb->mac_len;
}
\end{lstlisting}

The solution proposed in this paper leverages the existing cache system in the \lwt implementation of \ioam, \sr, and \rpl protocols, by fetching and using the cache earlier (\ie, before adding the new header). As a reminder, a \lwt encapsulation is attached to a route. Therefore, before a packet matches such a route, the cache, which is supposed to contain the corresponding destination entry (\dfn{dst}), and so the corresponding output device, is empty. Once a first packet matches the route, the \lwt encapsulation is applied to the packet, and the new destination entry is resolved and stored in the cache. After that, the cache is directly used every time another packet matches the same route, without resolving the new destination entry after the \lwt encapsulation anymore.

Listing~\ref{lst:pattern2} shows the simplified code pattern found in listing~\ref{lst:pattern}, with the mitigation solution applied. Line~\ref{lst:pattern2:1} uses a new function named \dfn{dst\_dev\_overhead()}. The code of this new function is shown in listing~\ref{lst:ddo}. Its purpose is to return the headroom size required by the output device (for the Layer-2 header and any extra data needed by the NIC) when the cache is not empty. Otherwise, if the cache is empty, the function returns the generic \dfn{mac\_len}. Therefore, the cache is now checked before the \lwt encapsulation. The logic behind lines~\ref{lst:pattern2:2}, \ref{lst:pattern2:3} and \ref{lst:pattern2:4} is the same as previously.

As a result, the first packet that matches a route with an attached \lwt encapsulation would see an empty cache. Should all specific conditions be met for the double-reallocation to happen, the issue would persist for that first packet only. Indeed, it is impossible to avoid it in that case because the cache is needed, which is not possible for the very first packet. On the other hand, subsequent packets are not impacted by the double-reallocation anymore, thanks to the proposed mitigation solution that leverages the cache.

\section{Performance Results}\label{perf}
\begin{figure*}[!t]
  \begin{center}
    \subfloat[\ioam]{
      \label{fig:ioam}
      \includegraphics[width=\columnwidth]{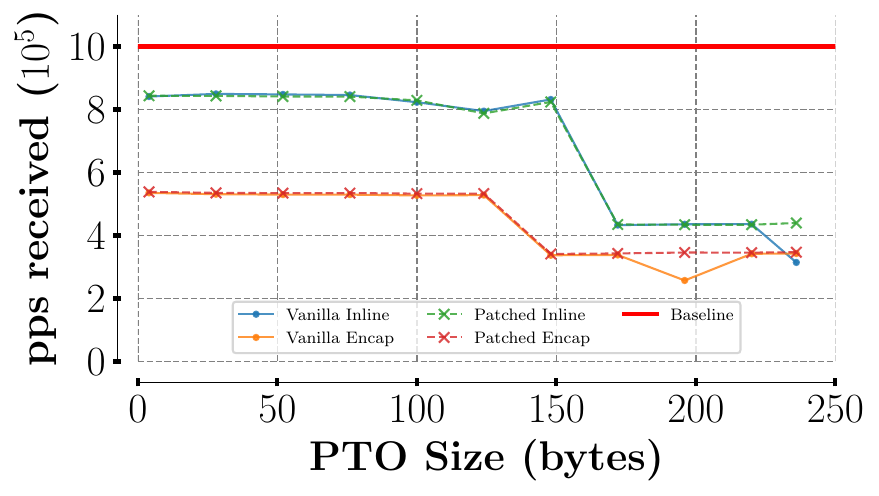}
    }
    \subfloat[\sr]{
      \label{fig:sr1}
      \includegraphics[width=\columnwidth]{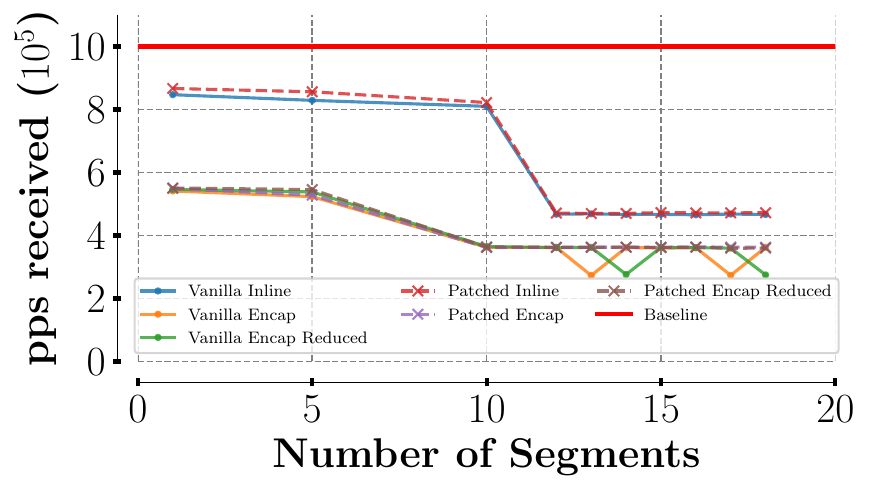}
    }

    \subfloat[\sr (L2 mode)]{
      \label{fig:sr2}
      \includegraphics[width=\columnwidth]{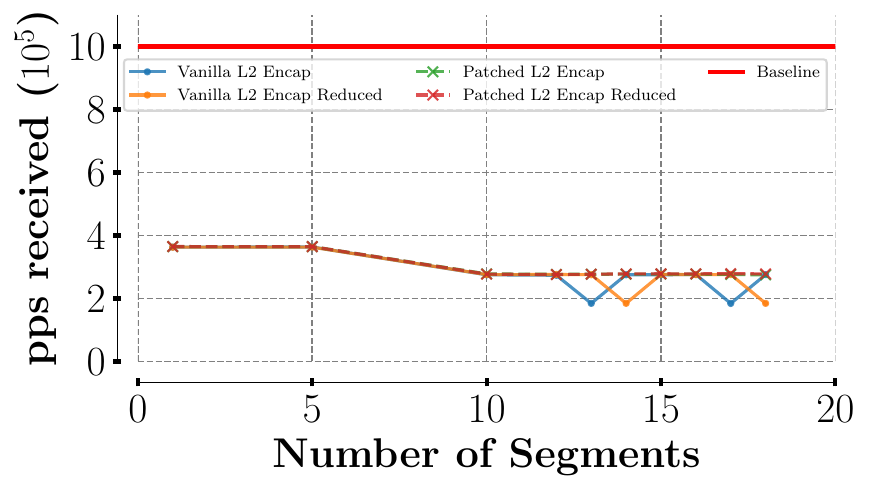}
    }
    \subfloat[\rpl]{
      \label{fig:rpl}
      \includegraphics[width=\columnwidth]{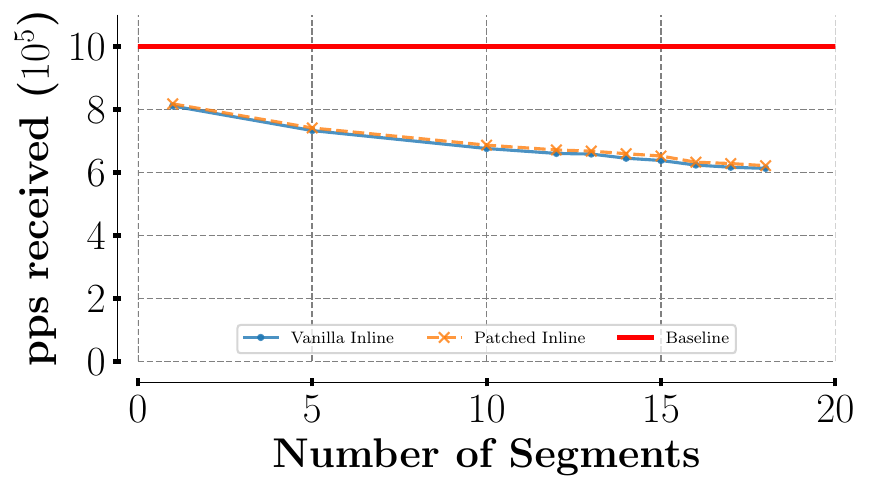}
    }
  \end{center}
  \caption{Performance evaluation.}
  \label{perf:fig}
\end{figure*}

\subsection{Methodology}
Performance evaluation follows the same methodology across all experiments.
We use Trex~\cite{trex}, a high-performance traffic generator, to transmit at line rate hand-crafted packets to the Device Under Test (\dut)\footnote{\dut specifications: \textit{x86 Intel(R) Xeon(R) CPU E5-2630 v3 @ 2.40GHz}, \textit{16 GB} RAM, and \textit{Intel XL710 Dual Port 40G QSFP+} NIC (i40e driver)}, which is a Linux machine with a recent kernel\footnote{Linux kernel version 6.12.} configured to forward traffic back to the generator in a port-to-port setup.

The key performance metric is the maximum number of packets per second (pps) the \dut can forward. We compare performance of a vanilla Linux kernel with our patched version. The \dut baseline (\ipsix packet forwarding rate) on a vanilla kernel is 1,000,000 packets per second, as indicated by the red lines in Fig.~\ref{perf:fig}. For each value on the plots, $5$ experiments are performed. Confidence intervals around the mean are computed but too tight to appear in the plots.

Overall, this methodology is similar to the methodology~\cite{draft-benchmark-srv6} proposed by the \textsc{IETF} to evaluate \sr performance.

\subsection{Results}
Fig.~\ref{fig:ioam} shows the forwarding capabilities for increasing sizes of the \ioam \pto. Performance degradation due to the double-reallocation issue is clearly visible with a vanilla kernel: one can see a forwarding rate drop occurring for \ptos of $196$ bytes with the Encap mode, and for \ptos of $236$ bytes with the Inline mode.  On average, performance decreases by $27.1$\% with a vanilla kernel when the issue occurs. The issue is completely mitigated in the patched version.

Fig.~\ref{fig:sr1} and~\ref{fig:sr2} show the forwarding capabilities for increasing number of segments in the \sr header.  In particular, Fig.~\ref{fig:sr2} specifically shows performance of \sr in L2 mode. In both figures, the vanilla kernel exhibits noticeable performance drops at specific segment counts. These drops occur at segment counts of $13$ and $17$ for the Encap mode (both normal and L2 versions), and at segment counts of $14$ and $18$ for the reduced (``Red'') Encap mode (both normal and L2 versions).  On average, the vanilla kernel experiences a $28.8$\% reduction in performance when the issue occurs. The issue is completely mitigated in the patched version.

Fig.~\ref{fig:rpl} demonstrates that performance remains consistent across the vanilla and patched kernels for \rpl. This is because \rpl is unaffected by the double-reallocation issue, as explained previously.

As a summary, one can conclude that the patch proposed in this paper effectively mitigates the double-reallocation issue, without affecting performance for normal cases where the double-reallocation does not happen.

\section{Conclusion}\label{ccl}
In this paper, we investigated a performance issue in the Linux kernel \lwt implementation of \ioam, \sr, and \rpl protocols. The issue, which occurs under specific conditions, causes a double-reallocation of socket buffers instead of a single one, resulting in significant performance degradation of up to 28.8\%.

We proposed a patch to mitigate the double-reallocation issue and, through performance comparisons between a vanilla kernel and the patched version, we demonstrated that the observed performance degradation was effectively mitigated in \ioam and \sr. Although \rpl was not directly affected by the issue, its implementation shared similar code patterns. As a preventive measure, we applied the patch to \rpl as well, ensuring that potential future problems are avoided.

\section*{Source Code}
The patch proposed in this paper has been merged in the Linux kernel and can be seen online~\cite{patch-url}.

\section*{Acknowledgments}
This work has been supported by the CyberExcellence project, funded by
the Walloon Region, under number 2110186, and the Feder CyberGalaxia project.

\small{
\bibliographystyle{isea}
\bibliography{Bibliography}
}

\end{document}